**The Quantization Conditions in Curved Spacetime**

**and Uncertainty–driven Inflation**

Paul J. Camp[1]

John L. Safko[2]

Running head: Quantization and Inflation in Curved Space

[1] College of Computing, Georgia Institute of Technology, Atlanta, GA 30332

Telephone 404-894-7435

[2] Department of Physics University of South Carolina, Columbia, South Carolina 29208

**Abstract**

An alternative inflationary model is proposed predicated upon a consideration of the form of the uncertainty principle in a curved background spacetime. An argument is presented suggesting a possible curvature dependence in the correct commutator relations for a quantum field in a classical background which cannot be deduced by simply extrapolation from the flat spacetime theory. To assess the possible consequences of this dependence, we apply the idea to a scalar field in a closed Friedmann–Robertson–Walker background, using a simple model for the curvature dependence (along the way, a previous result obtained by Bunch (1980) for the adiabatically expanded wave function is corrected). The result is a time–dependent cosmological constant, producing a vast amount of inflation that is independent of the mass of the matter field or its effective potential. Furthermore, it is seen that the field modes are initially zero for all wavelengths and come into being as the universe evolves. In this sense, the universe creates its contents out of its own expansion. At the end of the process, the matter field is far from equilibrium and essentially reproduces the initial conditions for the New Inflationary Model.

## 1. Introduction

In the absence of a consistent and finite quantum field theory of gravity, it has been fashionable in recent years to treat gravity classically and every other field quantum mechanically. The hope, of course, has been to see through the back reaction of the matter field on the geometry some sort of quantum behavior in the gravitational field. One might thereby gain some insights into the nature of a quantum gravity and see through a fog some of the properties it might possess. The remarkable successes of inflationary models (Guth 1981, Linde 1982, 1986a) and the intriguing suggestion of a connection between gravity and thermodynamics (Hawking 1975) perhaps point the way, albeit murkily, to a synthesis of quantum mechanics and general relativity (it should be noted in the latter case, however, that the significance of this connection has recently been called into question (Pringle 1989).)

All of these attempts at doing quantum field theory in a curved background rest upon an unspoken assumption regarding the proper way of doing the quantum mechanics of particles in that background. In flat spacetime the particle commutator relation is

$$[x^{\mu}, p^{\nu}] = i\hbar\,\eta^{\mu\nu} . \tag{1.1}$$

In the continuum limit, this leads naturally to the scalar field commutator

$$[\phi(x^{\mu}), \pi(x^{\nu})] = i\hbar\,\delta(x^{\mu} - x^{\nu}) . \tag{1.2}$$

Everyone is familiar with the argument leading to this result as it appears in every relativistic quantum mechanics text so it will not be repeated here (see e.g. Itzykson and Zuber, 1980). It is generally believed (though spelled out explicitly in few places) that the proper way of translating this formalism to a curved background is to make the minimal substitution

$$\eta^{\mu\nu} \rightarrow g^{\mu\nu} \tag{1.3}$$

which leads in the end to the same scalar field commutator.

There would seem to be a potentially serious flaw in this argument. The particle commutator is simply the mathematical realization of the position–momentum uncertainty relation, this latter being the fundamental physical principle. In the light of the uncertainty relation, we are not permitted to think of particles as mathematical points but must consider them as being spread out over a volume (this problem is even more acute in the continuum limit where each field mode is defined as a momentum eigenstate on a substantial patch of the spacetime). This calls into question the minimal substitution (1.3) since it implicitly regards the quantum behavior of a particle to depend only on the value of the metric at the center of that volume and not on how the metric is changing across the volume. It might be argued, then, that the correct substitution would be

$$\eta^{\mu\nu} \rightarrow f(R)\, g^{\mu\nu} \tag{1.4}$$

where $f(R)$ is a function of one or another of the various curvature tensors, may be operator–valued, and goes to one in the flat spacetime limit. This would lead to the scalar field commutator

$$[\phi(x^\mu), \pi(x^\nu)] = i\hbar f(R)\, \delta(x^\mu - x^\nu)\ . \qquad (1.5)$$

This commutator makes a certain amount of sense. A field is, after all, an inherently nonlocal quantity and one might expect its quantum behavior in a curved background to depend on nonlocal geometric quantities such as curvature. Whichever commutator one wishes to view as the most fundamental, field or particle, the other must be obtainable as a limiting case and so a curvature dependence would seem inescapable. At any rate, the possibility of a curvature–dependence cannot be dismissed out of hand in an extrapolation from flat to curved spacetime.

As it happens, it is completely equivalent to absorb the factor $f(R)$ into the equations of motion. The modified particle commutator relation is

$$[x^\mu, p^\nu] = i\hbar f(R)\, g^{\mu\nu} \qquad (1.6)$$

which may be accomplished through modification of the momentum operator

$$p^\mu = -\, i\hbar f(R)\, \nabla^\mu\ . \qquad (1.7)$$

It might be questioned whether this operator is still the generator of spacetime translations. It is, as may easily be seen by the usual method of expanding a displaced wave function in a Taylor series:

$$\psi(x - \delta) = \psi(x) - \delta\,\psi'(x) + \frac{1}{2}\delta^2\psi''(x) + \ldots = \exp\left[-\delta\frac{d}{dx}\right]\psi(\mathbf{x})\,. \qquad (1.8)$$

The generator of spacetime translations is the derivative operator which, because of its form, is usually identified with the momentum operator by dividing out a factor of $\hbar$. We may still do this if we instead divide out a factor of $\hbar f$

$$-\delta\frac{d}{dx} \rightarrow i\,\delta\frac{p}{\hbar f}\,. \qquad (1.9)$$

We must begin with a relativistically covariant wave equation such as the Klein–Gordon equation for a scalar field (we will develop an action principle later, though you could logically start at either point). We obtain the modified form of this equation by the substitution of (1.7) for the momentum operator into the invariant length of the momentum four vector, allowing the resulting operator to act on $\phi$ (the correspondence principle):

$$p_\mu p^\mu = E^2 - |\mathbf{p}|^2 = -m^2 \rightarrow \Box\phi + \frac{1}{\hbar f}(\nabla_\mu f)(\nabla^\mu\phi) + \frac{m^2}{\hbar f^2}\phi = 0\,. \qquad (1.10)$$

For this differential equation to possess a Green function, it must be possible to write it in self adjoint form, in which the differential operator has the form

$$\mathcal{D} = \nabla^\mu(q(x)\nabla_\mu) + r(x) \qquad (1.11)$$

where $q$ and $r$ are differentiable functions. This is easily done for our modified Klein–Gordon equation, which acquires the form

$$\nabla_\mu(\hbar f\,\nabla^\mu)\,\phi + \frac{m^2}{\hbar f}\phi = 0\,. \qquad (1.12)$$

The modified Klein–Gordon equation (1.14) has a conserved current density which may easily be verified to be

$$J^{\mu} = \frac{\hbar f}{2i\text{m}} \left( \phi^{*} \nabla^{\mu} \phi - \phi \nabla^{\mu} \phi^{*} \right) \qquad (1.13)$$

leading to the scalar product formula

$$<\phi, \theta> = -i \int_{\Sigma} (-g_{\Sigma})^{1/2} f (\phi^{*} \overleftrightarrow{\partial}_{\mu} \theta) \, d\Sigma^{\mu} \qquad (1.14)$$

where $\Sigma$ is a spacelike surface of simultaneity and $d\Sigma^{\mu} = n^{\mu} d\Sigma$, $n^{\mu}$ being a unit timelike vector orthogonal to $\Sigma$. The definition of a Hermitian operator, in the light of this scalar product formula, must therefore be slightly adjusted to

$$<\phi, \mathcal{O}\, \theta> = [\phi, \mathcal{O} f \theta] = [\mathcal{O} f \phi, \theta] = <\mathcal{O}\phi, \theta> \qquad (1.15)$$

where [ , ] designates the standard scalar product formula (for the Klein–Gordon field, equation (1.14) without the factor of $f$ in the integrand). The factor of $f$ must always be grouped to the right of the operator $\mathcal{O}$ and this grouping must also be observed when taking the matrix elements of an operator. If one substitutes equation (1.7) for the modified momentum operator into this expression for the scalar product, it is easy to verify by integration by parts that it is Hermitian.

This paper does not purport to be a complete exegesis of the above ideas. Rather, it should be regarded as a preliminary study designed to ascertain whether the modified quantum field theory is internally and externally consistent and to discover at least qualitatively the sort of effects which might be expected to follow from it. The model which we choose to study is dictated by these requirements and by the necessity that it be analytically tractable (though we do choose a model that is at least qualitatively reasonable). There are many investigations to which the modified theory might be appropriate. A study of the naked singularity problem at the endpoint of black hole radiation is a possibility. Another, and the one which we pursue, is a cosmological investigation of the consequences of a modified field theory in the very early universe. We obtain an inflationary theory which can use *any* scalar field (and very likely any field at all, scalar or not) as the inflaton, and may have any coupling to other fields as well. Along the way a few surprises will result such as the automatic vanishing of the cosmological constant.

A word on notation: Units are G = $\hbar$ = c = 1 unless otherwise stated or clearly implied by their presence in the equations. The rubrics SQFT and MQFT refer to, respectively, standard quantum field theory (without $f$) and modified quantum field theory (with $f$). We use the (− − −) sign convention in the terminology of Misner et al (1973), with metric signature being (+ − − −).

## 2. Mode Solutions for the Klein–Gordon Equation in a Closed Radiation Dominated Friedmann–Robertson–Walker Metric

We choose this background model as one which is thought to fairly well represent the early universe. The line element, expressed in terms of the conformal time parameter, is

$$ds^2 = C(\eta)\left[ d\eta^2 - \frac{dr^2}{1 - kr^2} - r^2\, d\theta^2 - r^2 \sin^2\theta\, d\phi^2 \right] \tag{2.1}$$

where

$$C^{1/2}(\eta) = a^* \sin \eta$$

$$0 \le \eta \le \pi \text{ (beginning to end of universe)}$$

$$a^{*\,2} = \frac{8\pi\rho_{ro} a_o^4}{3}$$

$$a_o = \text{current value.} \tag{2.2}$$

What to choose for *f*? There are a number of possibilities. It would seem likely that under extreme conditions the universe would if anything become more, rather than less, quantum mechanical and so it would make sense to choose a monotonically increasing scalar function of the curvature. Possibilities include functions of R, $R_{\mu\nu}R^{\mu\nu}$, or $R_{\mu\nu\lambda\rho}R^{\mu\nu\lambda\rho}$. The first of these is undesirable since it vanishes in empty space in a Friedmann–Robertson–Walker metric. The other two are dependent on fourth or higher powers of the time and end up yielding field equations that cannot be solved. We face the choice, then, of doing a numerical solution of a realistic *f* or to compromise by making an analytically tractable choice that at least qualitatively reflects the behavior of a realistic *f*. This compromise choice would make it easier to see what was going on in the theory and it seems logical as a first step. Our choice should then have only a time dependence (in our coordinate system), as do the other choices, but of a lower power to make solution of the field equations feasible. We will use

$$f = \exp\left[-\frac{R_{\mu\nu}n^{\mu}n^{\nu}}{R_o}\right] \tag{2.3}$$

where $n^{\mu}$ is a unit timelike Killing vector and the minus sign arises due to our choice of metric signature. This expression is not generally covariant but that is important only to the level that the results depend on our choice of Killing vector. Basically, all we are requiring with this choice is the existence of a timelike vector field at all points and times. As our goal is a qualitative assessment of the features of the theory, this seems like a reasonable compromise, at least insofar as it provides a reasonable behavior for $f$ and will allow us to work out the analytical bugs and see what sort of results should fall out of a theory such as this one. Later work can then probe more closely into possible $f$s that are more likely.

The constant $R_0$ is undetermined but must have units of curvature (i.e. $[\text{distance}]^{-2}$). The only natural constant hanging around out of which to construct $R_o$ is the Planck length

$$R_o = \ell_p^{-2} \tag{2.4}$$

and this is what we shall usually assume. It happens that there are no current cosmological observations which can be used to pin down $R_o$ with any precision. At any rate, our choice for $R_o$ necessarily implies that we need only investigate very early times in the history of the universe.

The Klein–Gordon equation, and its associated Lagrangian, are

$$\Box\phi + \frac{1}{f}(\nabla^{\mu}f)(\nabla_{\mu}\phi) + \frac{m^2}{f^2}\phi = 0$$

$$\mathcal{L} = \frac{1}{2}(-g)^{1/2}\left[f g^{\mu\nu}\phi_{;\mu}\phi_{;\nu} - \frac{m^2}{f}\phi^2\right] \qquad (2.5)$$

where the middle, friction–like coupling between matter field and curvature means that the matter energy tensor in isolation is not conserved but as we shall see enables a wholesale transfer of energy between the matter and gravitational fields. This will give rise to a variety of fascinating consequences. For our choice of metric and $f$, only the time derivative survives in this term. The timelike component of the Ricci tensor in our model is

$$R_{00} = -\frac{3}{\sin^2\eta} \ . \qquad (2.6)$$

The modified Klein–Gordon equation for field normal modes $w$ in general cannot be solved. It is, however, soluble in the limit of very early times when it becomes

$$\Box\, w - \frac{1}{CR_o}\frac{6}{h^3}w_{,0} + m^2\, w \exp\left[-\frac{6}{R_o\,\eta^2}\right] = 0 \quad \text{where} \quad \phi = \sum_{\lambda}(a_\lambda w_\lambda + a_\lambda^{\dagger} w_\lambda^{*}) \ . \qquad (2.7)$$

This equation may be solved by separation of variables $w = Z(\mathbf{x})\, T(\eta)$. The spatial part is

$$Z(\mathbf{x}) = \sin^{\ell}\chi\, C^{\ell+1}_{n-\ell}(\cos\chi)\, Y_{\ell}^{m}(\theta,\phi) \qquad (2.8)$$

where the Y's are spherical harmonics, the C's are Gegenbauer or ultraspherical polynomials, and the change of radial coordinate

$$r = \sin \chi \tag{2.9}$$

has been made. This part of the solution is of little interest for our current purposes as we seek information about the time evolution of the universe.

The time part may also be solved fairly easily. First of all, change the independent variable to

$$u = \frac{1}{\eta} \tag{2.10}$$

and the time part of the field equation satisfies

$$T'' + \alpha u T' + \left(\frac{k}{u^4} + \frac{\mu^2}{u^6}\exp(-\alpha u^2)\right) T = 0$$

$$\text{with} \quad \alpha = \frac{6}{R_o},\ \mu = ma^*,\ \text{and } k = n(n+2)\ . \tag{2.11}$$

At early times (large u) the last term in this equation may be neglected and it becomes soluble, after factoring out the SQFT solution, in terms of an error function:

$$T(\eta) = e^{-ik\eta}\left[\mathrm{erf}\left(\frac{b}{\eta}\right) - 1\right] b = \sqrt{\frac{a}{2}} \tag{2.12}$$

This solution possesses a number of intriguing properties, not the least of which is that it vanishes at time zero for all field modes. While the exact expression for this solution depends, of course, on the form chosen for $f(R)$, its qualitative behavior does not. So long as $f$ is taken to be a monotonically increasing function of the curvature, T will vanish at early times. The matter field has only a potential existence at the beginning of the universe.

The growth of T from nothing to something poses a bit of an interpretational difficulty as well since it obviously implies that probability is not independently conserved in the matter field. This means that the wave function is not generally normalizable. $\langle\phi|\phi\rangle$ is found to be time dependent and so cannot be absorbed into the definition of $\phi$. This fact lies at the root of the nonconservation of the matter energy tensor which will be described later and is obviously due to the nonlinear coupling between matter and gravitational fields. Again, it does not seem to depend in detail on the nature of *f* and embodies a sort of "transference of probability" from one field to the other. The matter field grows at the expense of the gravitational field and it is only a combination of the two that satisfies a conservation principle.

**3 Regularization and Renormalization in the Effective Action**

The presentation given here largely follows the standard development (see e.g. Birrell and Davies 1982 for a summary) with a few deviations caused by the presence of *f*. Consequently, we shall not spend a great deal of time on the formal manipulations but content ourselves with giving a broad outline of the procedure and indicating the major differences.

The basic idea is to find the back reaction of the quantum matter field on the gravitational field by replacing the classical energy tensor in the Einstein field equation with the expectation value of the corresponding quantum operator.

$$R_{\mu\nu} - \frac{1}{2} R\, g_{\mu\nu} + \Lambda_B\, g_{\mu\nu} = -\, 8\pi G_B <T_{\mu\nu}> \quad . \tag{3.1}$$

$<T_{\mu\nu}>$ will, of course, prove to be infinite as are most of the interesting quantities in quantum field theory. However, it shall be possible to absorb these infinities into the definitions of the gravitational coupling constants $\Lambda_{\Box}$ and $G_{\Box}$, along with a couple of others to be introduced later. The devil, as usual, is in the details.

$T_{\mu\nu}$ can be obtained from variation of the matter action $S_m$ with respect to the metric. We seek an effective action W which yields the expectation value $<T_{\mu\nu}>$ in the same way

$$<T_{\mu\nu}> = -\frac{2}{(-g)^{1/2}} \frac{\delta W}{\delta g^{\mu\nu}} \quad . \tag{3.2}$$

As it happens, this is ultimately given in the usual fashion by the Feynman propagator

$$W = -\left(\frac{i}{2}\right) \mathrm{Tr}\, [\, \ln(-G_F)\, ] = \int (-g(x))^{1/2}\, L_{eff}\, d^n x \tag{3.3}$$

$$L_{eff} = \frac{i}{2} \lim_{x \to x'} \int_{m^2}^{\infty} \frac{1}{f(x)}\, G_F(x,x')\, dm^2 \, . \tag{3.4}$$

Obtaining an expression for $G_F$ is enormously involved and could be the subject of a paper in itself – which in fact it has been. Interested readers should try to reproduce the local momentum space derivation not given by Bunch and Parker (1979) in their paper on the subject. The algebra is hair–raising. In terms of a proper time integral, the result is

$$G_F\,(x,x') = -\frac{i\,\Delta^{1/2}(x,x')}{(4\pi)^{n/2}} \int_0^{\infty} i\,ds\,(is)^{-n/2} \exp\left[ -\frac{ism^2}{f} + \frac{\sigma}{2isf} \right] F(x,x';is) \qquad (3.5)$$

where σ is half the geodesic distance from x to x', Δ is the Van Vleck–Morette determinant, n is the dimensionality of the spacetime, and

$$F(x,x';\,is) = a_{-1}(x,x')\,(is)^{-1} + a_o(x,x') + a_1(x,x')\,(is) + a_2(x,x')\,(is)^2\,. \qquad (3.6)$$

The a's are given by

$$a_{-1} = -\frac{1}{4}\,\sigma^{;\alpha} f_{;\alpha} \frac{\sigma}{f^2} - \frac{1}{12} f_{;\alpha\beta}\sigma^{;\alpha}\sigma^{;\beta} \frac{\sigma}{f^2}$$

$$a_o = 1 - \frac{1}{4}\frac{f_{;\alpha}}{f}\,\sigma^{;\alpha} - \frac{1}{6}\,i\,\frac{f_{;\alpha\beta}}{f}\,\sigma^{;\alpha}\sigma^{;\beta} - \frac{1}{12}\frac{f_{;\alpha\beta}}{f}\,\sigma_{;\alpha}\sigma^{;\beta} + \frac{1}{12}\,\sigma\frac{f_{;\lambda}{}^{\lambda}}{f^2}$$

$$a_1 = \frac{f}{6}\,R - \frac{1}{2}\,\sigma^{;\alpha}\left( \frac{f}{6}\,R_{;\alpha} + \frac{m^2 f_{;\alpha}}{f^2} \right) - \frac{1}{3}\,a_{\alpha\beta}\,\sigma^{;\alpha}\,\sigma^{;\beta} + \frac{1}{12}\frac{f_{;\lambda}{}^{\lambda}}{f}$$

$$a_2 = \frac{1}{2}\left( \frac{f}{6}\,R \right)^2 + \frac{1}{30}\,\Box R + \frac{m^2 f_{;\lambda}{}^{\lambda}}{6f^2} - \frac{f}{180}\,R^{\alpha\beta}R_{\alpha\beta} + \frac{f}{180}\,R^{\alpha\beta\gamma\delta}R_{\alpha\beta\gamma\delta}$$

$$a_{\alpha\beta} = \frac{f}{12} R_{;\alpha\beta} + \frac{m^2 f_{;\alpha\beta}}{2f^2}$$

$$-f\left( -\frac{1}{30} R_\alpha{}^\lambda R_{\lambda\beta} + \frac{1}{60} R^\kappa{}_\alpha{}^\lambda{}_\beta R_{\kappa\lambda} + \frac{1}{60} R^{\lambda\mu\kappa}{}_\alpha R_{\lambda\mu\kappa\beta} - \frac{1}{120} R_{;\alpha\beta} + \frac{1}{40} \Box R_{\alpha\beta} \right). \quad (3.7)$$

The major difference between this and the usual result, aside from the factors of $f$ scattered everywhere, is the presence of the $a_{-1}$ term which is not present in the usual expansion. This would seem at first glance to give us a divergence at the lower end of the proper time integral (3.5). However, this expansion is only valid for x in an infinitesimal neighborhood of x' and at the end of the renormalization we shall take the coincidence limit $x \to x'$. In this limit, $a_{-1}$ and all its relevant derivatives vanish so the integral is finite. It remains simply to determine which terms diverge in the coincidence limit $x \to x'$ and absorb them into an appropriate gravitational coupling constant. This means, in 4 dimensions, the terms $a_o$, $a_1$ and $a_2$. We can regularize the expression for $L_{eff}$ by analytically continuing the dimensionality n into the complex plane where, so long as $n \neq 4$ all terms are finite. The coincidence limit of the a's may then be taken. The divergent part of the effective Lagrangian is then

$$L_{div} = -\frac{1}{2(4\pi)^{n/2}}\left[\frac{1}{n-4} + \frac{1}{2}\left(\gamma + \ln\left(\frac{m^2}{f\kappa^2}\right)\right)\right]\left[\left(\frac{m^2}{f}\right)^2 \frac{4a_o}{n(n-2)} - \left(\frac{m^2}{f}\right)\frac{2a_1}{n-2} + a_2\right]$$

(3.8)

where $\gamma$ = Euler's constant and results from the expansion of a gamma function to order $n - 4$ so as to capture all the potentially divergent terms. The $\kappa$ is a mass rescaling factor intended to keep the units of $L_{eff}$ the same regardless of n; $\kappa \to 1$ as $n \to 4$.

Part or all of each of these terms may be absorbed into an appropriately modified bare gravitational Lagrangian

$$L_g^{bare} = \frac{1}{16\pi G_B}(R - 2\Lambda_B) \to -\left( A + \frac{\Lambda_B}{8\pi G_B} \right) + \left( B + \frac{1}{16\pi G_B} \right) R + C \,. \qquad (3.9)$$

By collecting like powers of the curvature out of the $a_i$ (A contains zeroth power, B the first power and C the second), we can obtain the renormalized gravitational coupling constants. The presence of terms of second order in the curvature (C) means that we must add quadratic terms to the bare gravitational Lagrangian to cancel them out. This is the case even in SQFT but there are rather strict observational limits on the magnitude of these terms (Stelle 1977, 1978; Horowitz and Wald 1978) and they can presumably be renormalized to zero. Similarly, B can be absorbed into the gravitational constant G. Again, there is no difference with the usual development other than some spacetime dependence in the divergent factor being removed. Λ, however, presents an interesting problem.

The renormalized cosmological constant is given by

$$\Lambda = \Lambda_B + \frac{32\pi m^4 G_B}{(4\pi)^{n/2}\, n(n-2)} \frac{1}{f^2} \left\{ \frac{1}{n-4} + \frac{1}{2}\left[ \gamma + \ln\left( \frac{m^2}{f\kappa^2} \right) \right] \right\} . \qquad (3.10)$$

There are, in fact, two ways to obtain a finite Λ from this expression. One is to assume that $\Lambda_B$ is a divergent function of spacetime that exactly cancels the second term in equation (3.10), leaving a finite result behind. The other is to take $\Lambda_B$ to be an infinite ( in the n=4 limit) *constant* which is just sufficiently large to cancel out the divergence but leave the spacetime dependence behind. This cannot be done with any of the other divergent terms.

One could argue that all three terms should be treated the same mathematically. On the other hand, $\Lambda$ is unique in that it has an intimate connection with the vacuum state energy of the matter field, from which the $f-$ dependent terms are arising. Discarding it entirely would imply that this connection is of little physical consequence whereas such phenomena as the Casimir effect point quite emphatically to the contrary. It is our opinion that renormalization is a mathematical procedure, not a physical imperative; there is no self evident "Law of Renormalization." For one thing, no one can tell us in advance when we would have to apply it. For some theories (not all), we can use this procedure to sweep embarrassing infinities under the rug long enough to extract meaningful predictions. Surely some day we shall have a theory which is finite from the beginning but until such time we will stick with the physics first. A physical argument has lead to the appearance of spacetime dependent terms in the gravitational coupling constants. Some of these terms are divergent. We will make the minimum adjustments necessary to remove the infinities (this is the only renormalization principle which makes sense to us) and leave as much of the spacetime dependence behind as we can. It would seem to us that anything more would be inconsistent with the physical importance of $\Lambda$.

Solve equation 3.10 for $\Lambda_B$ by evaluating the equation today, when $f = 1$. Let $\Lambda_o$ be today's value (presumably zero but in any case only a rescaling constant). Using the resulting value for $\Lambda_B$ in 3.10 and simplifying gives (in four dimensions)

$$\Lambda = \frac{3}{2} m^2 \left(1 - \frac{1}{f^2}\right). \tag{3.11}$$

The remaining linear and quadratic terms are removed in the usual way (see Birrell and Davies 1982 for details).

This is not quite the whole story, as may be seen by contemplating the energy tensor obtained from variation of the matter action

$$S = \int (-g)^{1/2} \frac{1}{2} \left( f g^{\alpha\beta} \phi_{;\alpha} \phi_{;\beta} - \frac{m^2}{f} \phi^2 \right) d^n x \tag{3.12}$$

which will give

$$T^{\mu\nu} = -\frac{1}{2} g^{\mu\nu} \left( f \phi_{;\alpha} \phi^{;\alpha} - \frac{m^2}{f} \phi^2 \right) + f \phi^{;\mu} \phi^{;\nu} - A^{\mu\nu\alpha\beta}{}_{;\alpha\beta} - A^{\alpha\beta\mu\nu}{}_{;\alpha\beta} + 2A^{\mu\alpha\nu\beta}{}_{;\alpha\beta} \tag{3.13}$$

where

$$A^{\alpha\beta\sigma\tau} = -\frac{f}{2R_o} \left( \phi_{;\gamma} \phi^{;\gamma} + \frac{m^2}{f} \phi^2 \right) n^\alpha n^\beta g^{\sigma\tau} . \tag{3.14}$$

The divergence of this energy tensor at early times is

$$T^{\mu\nu}{}_{;\nu} = -\frac{1}{2} f^{;\mu} \left( \phi_{;\alpha} \phi^{;\alpha} + \frac{m^2}{f} \phi^2 \right) \tag{3.15}$$

which is manifestly nonzero.

Non–conservation of $T_{\mu\nu}$ is not necessarily unexpected nor is it necessarily a bad thing. The Lagrangian 2.5 is formally analogous to that for a problem such as a rocket with a time varying mass. I would not in that case expect the energy of the rocket to be conserved for I would have neglected that energy carried off by the reaction products. Furthermore, we know that $T^{\mu\nu}$, as a consequence of Noether's theorem, is conserved only if the Lagrangian is translationally invariant. Since *f* is curvature–dependent, this translational invariance is broken. Lastly, recall our mode solutions 2.12 which are initially zero. How could something which does not initially exist but does later on possibly have a conserved $T^{\mu\nu}$? Evidently, energy is being pumped in wholesale from the gravitational field as the matter field is being created and only the sum of the two can possibly be conserved.

This is not in fact a new problem. DeWitt (1975) has pointed out that in any nonstationary spacetime it is impossible to define a $T^{\mu\nu}$ that is simultaneously normal ordered in both the "in" and "out" regions, has matrix elements which are smooth functions, and has a divergence which vanishes everywhere. According to DeWitt, the correct procedure is to give up the normal ordering and use a subtraction procedure that respects the conservation equation

$$T^{\mu\nu}{}_{;\nu} = 0 \ . \qquad (3.16)$$

This will require a modification of the standard subtraction procedure as we shall have to show that the nonconserved part of $T^{\mu\nu}$ can be absorbed into the renormalized coupling constants along with the infinite parts. We will postpone the demonstration of this until the next section as it requires an expression for $\phi$ in terms of the curvature which will be given by the adiabatic expansion. However, we should point out now that carrying out DeWitt's prescription will require the addition of a geometric counterterm, and one more coupling constant, in the Einstein equation. Define the auxiliary tensor

$$\tau^{\mu\nu}{}_{;\nu} = T^{\mu\nu}{}_{;\nu} \qquad (3.17)$$

as the nonconserved part of $T^{\mu\nu}$ (obviously there is some gauge ambiguity in this definition). The first three terms of $\tau^{\mu\nu}$ contribute to the renormalized form of $\Lambda$, G and the quadratic coupling constant mentioned above. The remaining terms in the expansion of $\tau^{\mu\nu}$ require the additional counterterm. There is no contribution to this extra counterterm from the conserved part of $<T^{\mu\nu}>$ as it has no divergence factors higher than quadratic order in the curvature. Hence, the new counterterm and its coupling constant are determined entirely by $\tau^{\mu\nu}$. Unfortunately, we cannot give a closed expression for this counterterm as we shall have at our disposal only an iterative expansion of the wave function to employ in evaluating $\tau^{\mu\nu}$. Nevertheless, the terms are purely curvature dependent, justifying a counterterm on the gravitational side of the equation, and a finite number of iterations serves to determine the new coupling constant since no other factors will make a contribution of higher order than quadratic.

The term $\tau^{\mu\nu}$ also has lower order terms and these will make contributions to the usual constants, especially $\Lambda$. We can make two arguments here. For one, if it is possible to retain the spacetime dependence in $\Lambda$ then we should for the same reasons as given before. For another, unless we remove *all* of $\tau^{\mu\nu}$ we have not truly followed DeWitt's prescription and have left some of the nonconserved part behind. On Mondays, Wednesdays and Fridays we believe one of these. On Tuesdays, Thursdays and Saturdays we believe the other. On Sundays we watch the birds. To allow this paper to be read throughout the week, we shall include a parameter $\lambda$ that can be set equal to one or zero (or if you tilt your head a bit, viewed as a bird) depending as the reader wishes to entirely remove $\tau^{\mu\nu}$ or not.

Using the mode solutions given in section 2 above, equation 3.17 can be approximately integrated at early times

$$\tau^{\mu\nu} \approx \frac{\alpha}{2\eta^6} \exp\left[\frac{\alpha}{2\eta^2}\right] g^{\mu\nu} \phi^2 \tag{3.18}$$

which is derivable from the action

$$S_\tau = \int (-g)^{1/2} \left[ \frac{\alpha}{2\eta^6} \exp\left[\frac{\alpha}{2\eta^2}\right] \phi^2 \right] d^4x \, . \tag{3.19}$$

The semiclassical Einstein equation is as before except that instead of $\langle T^{\mu\nu} \rangle$ on the right hand side, we have $\langle T^{\mu\nu} - \tau^{\mu\nu} \rangle$. The renormalization procedure is essentially unchanged, simply including the effective Lagrangian for $\tau^{\mu\nu}$

$$L_\tau = i \frac{\alpha}{2\eta^6} \exp\left[\frac{\alpha}{2\eta^2}\right] \lim_{x \to x'} G_F(x,x') \tag{3.20}$$

which has divergent part

$$L_\tau^{div} = -\frac{\alpha}{(4\pi)^{n/2}\,\eta^6}\exp\left[\frac{\alpha}{2\eta^2}\right]\left[\frac{1}{n-4}+\frac{1}{2}\left(\gamma+\ln\left(\frac{m^2}{f\kappa^2}\right)\right)\right]\left[\frac{m^2}{f}\frac{2a_o}{2-n}+a_1\right] \tag{3.21}$$

From here it is simply a matter of churning through the algebra to find the new expression for $\Lambda$

$$\Lambda = \Lambda_o + \frac{3}{2}m^4\,\frac{1-\frac{1}{f^2}-\lambda\frac{2\alpha}{m^2}\left(\frac{1}{\eta_o^6}\exp\left[\frac{\alpha}{2\eta_o^2}\right]-\frac{1}{f\eta^6}\exp\left[\frac{\alpha}{2\eta^2}\right]\right)}{m^2+\lambda\alpha f\frac{1}{\eta^6}\exp\left[\frac{\alpha}{2\eta^2}\right]} \tag{3.22}$$

in four dimensions with the subscript "o" denoting today's values and $\lambda$ being the aforementioned constant which may be set equal to 0 or 1 depending on how much of $\tau^{\mu\nu}$ one wishes to banish. In the former case, the previous expression, 3.11, is recovered.

If $\lambda = 0$, then $\Lambda$ starts at a very large value ($3m^2/2$) and decreases smoothly to $\Lambda_o$. If $\lambda = 1$, then at about the Planck time $\Lambda$ rises from $\Lambda_o$ to a very large value and then decreases smoothly to $\Lambda_o$ again. In other words, there isn't a great deal of practical difference between the choices. However, note that in previous inflationary models, the current smallness of $\Lambda$ has been a bit of a puzzle. Here, it is a necessary consequence of the theory. The vacuum energy density itself will also be seen to vanish.

Though we now have expressions for the renormalized coupling constants, we do not unfortunately know enough about the renormalized Lagrangian to functionally differentiate it and obtain an energy tensor. The reason is that we only have an asymptotic expansion for the Feynman propagator and that is simply not accurate enough. The energy tensor must be attacked directly.

### 4. Regularization and Renormalization in the Energy Tensor

There are a variety of methods for systematically isolating the finite parts of $T^{\mu\nu}$ directly. The method which we shall employ, which is especially suited to the Friedmann–Robertson–Walker metric, is known as adiabatic regularization (Parker and Fulling 1974, Fulling and Parker 1974, Fulling et al 1974). Equivalence to renormalization of the gravitational coupling constants by dimensional regularization was demonstrated by Bunch (1980). As discussed in these papers, the term "adiabatic order" refers to the number of derivatives of the metric involved. In the following, we will use the notation

$$D = \frac{\dot{C}}{C}, \qquad F = \frac{\dot{f}}{f} \quad \text{and} \quad \Omega_k^2 = k^2 + \frac{Cm^2}{f^2}. \tag{4.1}$$

The adiabatic expansion of the wave function is essentially a WKB expansion in the time part of the differential equation, which of course assumes $T(\eta)$ is slowly varying. We must therefore factor out the error function part and perform an expansion of the remaining bits

$$T = \left[ \operatorname{erf}\left( \sqrt{\frac{\alpha}{2}} \frac{1}{\eta} \right) - 1 \right] C^{-1/2} \chi \tag{4.2}$$

Call the term in square brackets $T_o$ . Making this change, the time part of the equation of motion becomes

$$\ddot{\chi} - F\,\dot{\chi} + (\Omega_k^2 + \frac{1}{2}DF - \frac{1}{2}\dot{D} - \frac{1}{4}D^2)\,\chi = 0\,. \tag{4.3}$$

It appears, incidentally, that Bunch (1980) has made a minor error at this point. While it is true that

$$\frac{1}{2}\dot{D} + \frac{1}{4}D^2 \approx 0$$

at early times, this is not true in general and the term should not therefore be omitted as he has done. Perform the WKB expansion on this equation

$$\chi = \frac{1}{(2W)^{1/2}}\exp\left[-i\int^{\eta} W(\eta')d\eta'\right] \tag{4.4}$$

and find that W satisfies

$$-\frac{\ddot{W}}{2W} + \frac{3}{4}\frac{\dot{W}^2}{W^2} - W^2 + F\frac{\dot{W}}{2W} + iFW + \Omega_k^2 + \frac{1}{2}DF - \frac{1}{2}\dot{D} - \frac{1}{4}D^2 = 0\,. \tag{4.5}$$

This may be written as a recursion relation for W

$$W_{n+1}^2 = -\frac{\ddot{W}_n}{2W_n} + \frac{3}{4}\frac{\dot{W}_n^2}{W_n^2} + F\left(\frac{\dot{W}_n}{2W_n} + i\,W_n + \frac{1}{2}D\right) + \Omega_k^2 - \frac{1}{2}\dot{D} - \frac{1}{4}D^2\,. \tag{4.6}$$

It is not difficult, though it is enormously tedious, to iterate the solution of this equation to fourth order. It is understandable, then, that Bunch (1980) would have made some algebraic errors in the SQFT version. We had the advantage over him in the subsequent development of symbolic manipulation computer programs and this iteration has been performed both by hand and by MAPLE. The corrected result is

$$W = \Omega_k - \frac{1}{8}\frac{Cm^2}{\Omega_k^3 f^2}\left(D^2 + \dot{D}\right) + \frac{5}{32}\frac{C^2m^4}{\Omega_k^5 f^4}D^2$$

$$+ \frac{1}{32}\frac{Cm^2}{\Omega_k^5 f^2}\left(\dddot{D} + 6D^2\dot{D} + 4D\ddot{D} + D^4 + 3\dot{D}^2\right)$$

$$- \frac{1}{128}\frac{C^2m^4}{\Omega_k^7 f^4}\left(28D\ddot{D} + 19\dot{D}^2 + 122D^2\dot{D} + 47D^4\right) + \frac{221}{256}\frac{C^3 m^6}{\Omega_k^9 f^6}\left(\dot{D}D^2 + D^4\right)$$

$$\underline{- \frac{1087}{2048}\frac{C^4 m^8}{\Omega_k^{11} f^8}D^4} - \frac{1}{128\,\Omega_k^3}\boxed{\left(D^4 - 8\dddot{D} - 4\dot{D}^2 + 4\dot{D}D^2 - 8D\ddot{D}\right)}$$

$$- \frac{1}{8\Omega_k}\left(\boxed{D^2 + 2\dot{D}} - 2DF\right) + \frac{1}{2}\frac{Cm^2}{\Omega_k^3 f^2}\left(4DF + 3\dot{F}\right) - \frac{5}{8}\frac{C^2m^4}{\Omega_k^5 f^4}DF + \frac{1}{2}iF$$

$$- \frac{1}{64}\frac{Cm^2}{\Omega_k^5 f^2}\boxed{\left(19D^2\dot{D} - 6D\ddot{D} + 3D^4 + 6\dot{D}^2\right)} + \frac{25}{256}\frac{C^2m^4}{\Omega_k^7 f^4}\boxed{\left(D^4 + 2D^2\dot{D}\right)}. \tag{4.7}$$

The boxes set off terms arising from the factors previously neglected by Bunch (see footnote to equation 4.3). The underlined term differs from that given by Bunch by a numerical factor and appears to have been an algebraic error. The corrected version of Bunch's SQFT result may be obtained from this one by setting $f = 1$.

We can now resolve the discussion that was hinted at in the paragraphs following equation 3.17 regarding the subtraction of $\tau^{\mu\nu}$. Recall that the subtraction procedure we utilize involves the removal of finite terms in $\tau^{\mu\nu}$ corresponding to the nonconserved part of the energy tensor. $\tau^{\mu\nu}$ is proportional to $\phi^2$ and therefore contains only even adiabatic orders. Both $<T^{\mu\nu} - \tau^{\mu\nu}>$ and $<\tau^{\mu\nu}>$ contribute to the zeroth, second and fourth order adiabatic terms in the Einstein equation, resulting in renormalization of, respectively, $\Lambda$, G and a third constant A associated with quadratic terms. Only $\tau^{\mu\nu}$ contributes to sixth and higher order terms, uniquely defining all the counterterms necessary to cancel it out, which are up to a constant simply those obtainable from continuation of the adiabatic expansion of $\phi$ and therefore are purely geometric. Furthermore, since only $\tau^{\mu\nu}$ contributes to sixth and higher order terms, only one additional coupling constant, call it B, need be used which is uniquely determined by the sixth order term in the expansion of $\tau^{\mu\nu}$ : B is whatever is necessary to eliminate the sixth order term. It will then necessarily eliminate all higher order terms as well.

Define the conserved energy tensor $t^{\mu\nu} = T^{\mu\nu} - \tau^{\mu\nu}$. Due to the symmetries of the Friedmann–Robertson–Walker cosmology, the driving term in Einstein's equation is $<t^{oo}>$ so we concentrate on its form (see Fulling et al 1974 for this expression)

$$<t^{oo}> = \frac{<\int d^3x\,(-g_\Sigma)^{1/2}\, t_{oo}>}{\int d^3x(-g_\Sigma)^{1/2}} = \frac{1}{2} <f\,(f_{;o})^2> - \frac{1}{2} < f\,\phi\nabla^2\phi - \frac{m^2}{f}\,\phi^2> - \frac{\alpha}{2} < \frac{fC}{\eta^6}\,\phi^2> . \tag{4.8}$$

Here, $\Sigma$ is a spacelike hypersurface on which expectation values are to be evaluated and we have employed 3.18 and 3.13 for $\tau^{\mu\nu}$ and $T^{\mu\nu}$ respectively.

Now we can expand $\phi$ in terms of the field modes previously obtained (section 2 above)

$$\phi = \sum_\lambda (a_\lambda w_\lambda + a_\lambda^\dagger w_\lambda^*) \tag{4.9}$$

where the raising and lowering operators, due to the absorption of $f$ into the definition of the canonical momentum, satisfy the usual commutator relations. To a good approximation, the mode sum can be replaced by an integral and shown to be essentially of the time part of the field modes (see Fulling et al 1974 again). Employing the equations of motion to simplify the integral, we ultimately find that the vacuum state expectation value is

$$<t_{oo}> = \frac{T_o^2}{4\pi^2 C}\,\mathcal{D}^2 \int d\mu(k)\ |\chi|^2\, e^{-\sigma_k} \tag{4.10}$$

where the exponential is the usual regulator function to make the integral finite (eventually we take the limit $\sigma \rightarrow 0$), the differential operator is defined by

$$\mathcal{D}^2 = \frac{f}{4}\,\partial_\eta^{\,2} - \frac{\dot{f}}{4}\,\partial_\eta + \frac{1}{4}\,D\dot{f} + f\left( \partial_\sigma^{\,2} + \frac{Cm^2}{f^2} \right) - \frac{\alpha f C}{2\eta^6} + \frac{m^2}{f} \tag{4.11}$$

and for the Friedmann–Robertson–Walker cosmology, $\int d\mu(k) \rightarrow \sum_{k} k^2$.

So far, no adiabatic approximation has been made. What we do now is evaluate this regularized expression using the exact wave functions, do the same with the adiabatic approximation to the wave functions, evaluate the difference between those two expressions, expand in the regulator parameter $\sigma$ (the divergent parts should cancel), and take the limits $\sigma, \eta \rightarrow 0$ to find the finite remainder. Sounds easy, takes about forever.

The manipulations are not particularly difficult but are quite time–consuming, especially for the adiabatic part. Essentially, we substitute the adiabatically expanded wave function 4.4–7 into 4.10 for $<t^{oo}>$, perform a great many integrals, and operate on the result with the operator $\mathcal{D}^2$. Along the way, we discard all terms of adiabatic order greater than four. The end result is, for the adiabatic $<t_{00}>_A$,

$$<t_{00}>_A = -\frac{T_0^2}{4\pi^2 C}$$

$$\times\Bigg\{ -\frac{Cm^2}{8f}(D-2F)^2\left(\ln\left(\frac{C^{1/2}m\sigma}{2f}\right)+\gamma+\frac{3}{2}\right) - \frac{Cm^2}{8f}(\dot{D}-2\dot{F})\left(\ln\left(\frac{C^{1/2}m\sigma}{2f}\right)+\gamma+1\right)$$

$$-\frac{f}{96}(\dot{D}^2+D\ddot{D}+\dddot{D}) + \frac{CFm^2}{8f}(D-2F)\left(\ln\left(\frac{C^{1/2}m\sigma}{2f}\right)+\gamma+1\right) + \frac{fF}{96}(D\dot{D}+\ddot{D})$$

$$+\frac{6F}{\sigma^4}+\frac{Cm^2}{2f\sigma^2}-\frac{3C^2m^4}{8f^3}\left(\ln\left(\frac{C^{1/2}m\sigma}{2f}\right)+\gamma+\frac{29}{12}\right)+\frac{Cm^2}{8f}(2D^2+\dot{D})$$

$$+\frac{f}{160}(\dddot{D} + D^2\dot{D} + 4D\ddot{D} + 3\dot{D}^2 + D^4)$$

$$-\frac{f}{2240}(28D\ddot{D} + 21\dot{D}^2 + 126D^2\dot{D} + 49D^4)$$

$$+\frac{847f}{81\,920}(D^2\dot{D} + D^4) - \frac{379f}{49\,280}D^4 + \left[\frac{f}{4\sigma^2} - \frac{3Cm^2}{f}\left(\ln\left(\frac{C^{1/2}m\sigma}{2f}\right) + \gamma + \frac{5}{6}\right)\right]DF$$

$$-\frac{Cm^2}{6f}\left(\ln\left(\frac{C^{1/2}m\sigma}{2f}\right) + \gamma + \frac{4}{3}\right)(4DF + 3\dot{F}) - \frac{Cm^2}{4f}DF$$

$$-\frac{f}{160}(19D^2\dot{D} - 6D\ddot{D} + 3D^4 + 6\dot{D}^2) + \left(\frac{Cm^2}{f^2} + \frac{afC}{2\eta^6} - \frac{m^2}{f} - \frac{1}{4}D/F\right)$$

$$\times\left[\frac{1}{\sigma^2} + \frac{Cm^2}{2f^2}\left(\ln\left(\frac{C^{1/2}m\sigma}{2f}\right) + \gamma + \frac{1}{2}\right) + \frac{1}{24}(D^2 + \dot{D}) - \frac{1}{48}D^2\right.$$

$$-\frac{f^2}{240Cm^2}(\dddot{D} + 6D^2\dot{D} + 4D\ddot{D} + 3\dot{D}^2 + D^4)$$

$$+\frac{f^2}{1680Cm^2}(28D\ddot{D} + 21\dot{D}^2 + 126D^2\dot{D} + 49D^4)$$

$$-\frac{2541}{12\,280}\frac{f^2}{Cm^2}(D^2\dot{D} + D^4) + \frac{379}{18\,480}\frac{f^2}{Cm^2}D^4 + \frac{1}{4}\left(\ln\left(\frac{C^{1/2}m\sigma}{2f}\right) + \gamma + 1\right)DF$$

$$\left.-\frac{1}{18}(4DF + 3\dot{F}) + \frac{1}{6}DF + \frac{f^2}{240\,Cm^2}(19D^2\dot{D} - 6D\ddot{D} + 3D^4 + 6\dot{D}^2)\right]\Bigg\}. \tag{4.12}$$

The exact solution of the wave equation 4.3 is

$$\chi = \frac{1}{\sqrt{\kappa}}\, e^{-i\,\kappa\,\eta} \quad \text{with} \quad \kappa^2 = k^2 + \frac{Cm^2}{f^2} + \frac{1}{2}DF - \frac{1}{2}\dot{D} - \frac{1}{4}D^2 \equiv k^2 + \delta^2 \qquad (4.13)$$

which, when inserted in equation 4.10, gives a properly regulated expression for $<t_{oo}>_{exact}$

$$<t_{oo}>_{exact} = \frac{T_o^{\,4}}{4\pi^2 C}\left\{ \frac{f}{4}\dot{\delta}^2\left( \ln\!\left(\frac{\delta\sigma}{2}\right) + \gamma + 2\right) - \frac{f}{4}\,\delta\ddot{\delta}\left( \ln\!\left(\frac{\delta\sigma}{2}\right) + \gamma + 1\right)\right.$$

$$- \frac{fF}{4}\,\delta\dot{\delta}\left( \ln\!\left(\frac{\delta\sigma}{2}\right) + \gamma + 1\right) - \frac{f\delta^2}{2\sigma^2} - \frac{3f\delta^4}{8}\left( \ln\!\left(\frac{\delta\sigma}{2}\right) + \gamma + \frac{11}{16}\right) - \frac{6f}{\sigma^4} + \frac{15f\delta^4}{32}$$

$$\left. + \left( \frac{\alpha f C}{2\eta^6} - \frac{m^2}{f} + \frac{Cm^2}{f^2} - \frac{1}{4}DfF\right)\left[ -\frac{\delta^2}{2}\left( \ln\!\left(\frac{\delta\sigma}{2}\right) + \gamma + 1\right) - \frac{1}{\sigma^2} + \frac{\delta^2}{4}\right]. \qquad (4.14)$$

When we use these two expressions for $<t_{oo}>$ to calculate the renormalized energy tensor

$$<t_{oo}>_{ren} = \lim_{\sigma\to 0}\left( <t_{oo}>_{exact} - <t_{oo}>_A \right) \qquad (4.15)$$

terms quartically, quadratically and logarithmically divergent for large $\sigma$ exactly cancel, but infrared divergences are introduced. These turn out to be artifacts of the subtraction procedure since the adiabatic expansion is designed to accurately describe ultraviolet divergences but is not uniformly asymptotic as $k \to 0$. Hence, the infrared divergences are spurious, arising out of the bad low frequency behavior of the adiabatic expansion (Fulling and Parker 1974). They may be safely discarded to yield the finite remainder

$$<t_{oo}>_{ren} = \frac{T_o^{\,2}}{4\pi^2 C}\left\{ -\frac{3f}{8}\left( \gamma + \frac{23}{16}\right)\left(\frac{DF}{2}\right)^4 + \frac{f}{96}\left( \dot{D}^2 + D\ddot{D} + \dddot{D}\right)\right.$$

$$-\frac{fF}{96}(D\dot{D}+\ddot{D})-\frac{f}{160}(\dddot{D}+D^2\dot{D}+4D\ddot{D}+3\dot{D}^2+D^4)$$

$$+\frac{f}{2240}(28D\ddot{D}+21\dot{D}^2+126D^2\dot{D}+49D^4)-\frac{847f}{81\,920}(D^2\dot{D}+D^4)$$

$$+\frac{379f}{49\,280}D^4+\boxed{\frac{f}{160}(19D^2\dot{D}-6D\ddot{D}+3D^4+6\dot{D}^2)}$$

$$+\left(\frac{\alpha fC}{2\eta^6}+\frac{Cm^2}{f^2}-\frac{m^2}{f}-\frac{f}{4}DF\right)$$

$$\times\left[-\frac{1}{24}DF-\frac{1}{2}(\gamma+1)DF-\frac{1}{48}(D^2+2\dot{D})+\frac{1}{18}(4DF+3\dot{F})\right]$$

$$+(1-f)\Bigg[\frac{1}{240}(\dddot{D}+6D^2\dot{D}+4D\ddot{D}+3\dot{D}^2+D^4)$$

$$-\frac{1}{1680}(28D\ddot{D}+21\dot{D}^2+126D^2\dot{D}+49D^4)+\frac{2541}{12\,280}(D^2\dot{D}+D^4)-\frac{379}{18\,480}D^4$$

$$-\boxed{\frac{1}{240}(19D^2\dot{D}-6D\ddot{D}+3D^4+6\dot{D}^2)}\Bigg]\Bigg\} \tag{4.16}$$

or, evaluated at early times,

$$\langle t_{oo}\rangle_{ren}=\frac{T_o^2}{4\pi^2 C}\left\{-\frac{3f}{8}\left(\gamma+\frac{23}{16}\right)\frac{\alpha^4}{\eta^{16}}+\frac{f}{48}\left(\frac{2}{\eta^4}-\frac{4}{\eta^3}-\frac{3}{\eta^2}\right)-\frac{f\alpha}{24}\left(\frac{1}{\eta^5}+\frac{1}{\eta^6}\right)\right.$$

$$-\frac{f}{80}\left(\frac{10}{\eta^4}-\frac{16}{\eta^3}-\frac{3}{\eta^2}\right)-\frac{f}{560}\left(\frac{56}{\eta^3}+\frac{35}{\eta^4}\right)-\frac{847}{10\,240}\frac{f}{\eta^4}+\frac{379}{3080}\frac{f}{\eta^4}$$

$$+\boxed{\frac{f}{10}\left(\frac{3}{\eta^3}-\frac{5}{\eta^4}\right)}+\left(\frac{\alpha fC}{2\eta^6}+\frac{Cm^2}{f^2}-\frac{m^2}{f}-\frac{f}{4}DF\right)\left[\frac{1}{12}\frac{1}{\eta^4}+(\gamma+1)\frac{\alpha}{\eta^4}+\frac{1}{18}\frac{\alpha}{\eta^4}\right]$$

$$+ (1 - f)\left[ -\frac{1}{120}\left(\frac{3}{\eta^2} + \frac{16}{\eta^3} + \frac{20}{\eta^4}\right) + \frac{1}{420}\left(\frac{56}{\eta^3} + \frac{35}{\eta^4}\right) + \frac{2541}{1535}\frac{1}{\eta^4} - \frac{379}{1155}\frac{1}{\eta^4} - \boxed{\frac{1}{15}\left(\frac{3}{\eta^3} - \frac{5}{\eta^4}\right)} \right]\Bigg\}. \qquad (4.17)$$

**5. Evolution of the Friedmann–Robertson–Walker Scale Factor**

The evolution equation for a(t) is

$$\dot{a}^2 = -1 + \frac{\Lambda}{3} a^2 + \frac{8\pi}{3}(\rho_c + \rho_q) a^2 \qquad (5.1)$$

where $\rho_c$ and $\rho_q$ represent the classical (background) and quantum (due to $\phi$) energy densities. It happens that each term in the quantum energy tensor 4.17 peaks rather sharply in the vicinity of the Planck time, mainly due to our choice of $R_o$, exhibiting near impulse–like behavior. While this may present a mechanism for singularity avoidance, it is occurring in a regime in which we should properly include the effects of quantum gravity anyway. The only term in 5.1 which survives significantly after the Planck time is the second, cosmological, term and we therefore focus our attention on it. If I begin at a few Planck times, it is quite large and is in fact by far the dominant term in the back reaction equation.

The expression 3.22 for $\Lambda$ is far too complicated to use in 5.1 but we can get an idea of what is going on by dividing the problem up into several time ranges in which various parts of $\Lambda$ dominate. This division is shown in Table I.

**Table I: Approximate forms and ranges of validity for $\Lambda$**

| $\Lambda$ | Range of $\eta$ ($t = (a^*/2)\,\eta^2$) |
| --- | --- |
| (a) $3m^2 \exp\left[-\frac{\alpha}{\eta^2}\right]$ | $\eta \lesssim \left(\frac{\alpha}{m^2}\right)^{1/6}$ |
| (b) $\frac{3\alpha}{\eta^6}$ | $\left(\frac{\alpha}{m^2}\right)^{1/6} \lesssim \eta \lesssim \left(\frac{2\alpha}{m^2}\right)^{1/6}$ |
| (c) $\frac{3}{2} m^2 \left(1 - \exp\left[-\frac{\alpha}{\eta^2}\right]\right)$ | $\eta \gtrsim \left(\frac{2\alpha}{m^2}\right)^{1/6}$ |

Case *a* is not especially useful since, again, it is not really valid much after the Planck time. Case *b* gives the inflationary solution

$$a(t) = a_o' \exp\left[-\frac{a^{*3}}{4}\sqrt{\frac{\alpha}{t}}\right] \qquad (5.2)$$

where $a_o'$ is set by the terminal value at the end of case *a*. Three important points should be noted about this solution. First, it is independent of the mass of the scalar field. This is as expected since in the scalar field wave equation, *f* appears in the denominator of the mass term so at sufficiently early times the field behaves as if it were massless. Furthermore, the same is true of any coupling term in the wave equation – at sufficiently early times, any coupling terms to other fields will be negligible in comparison to the derivative terms of the wave equation so the field behaves not only as if it were massless but as if it were free as well. Consequently, the inflationary solution 5.2 will follow from *any* scalar field. It seems likely to us that it will follow from any field at all, scalar or not, but this has not yet been proven.

Second, it inflates prodigiously. According to Fakir and Unruh (1990) and Brandenberger (1985), a minimum of 60 or so e–fold increase in size is required to solve all the problems (except the vanishing of $\Lambda$) which plague classical cosmology. Since $a^* \sim 10^{57}$ , we have certainly achieved that.

Third, though the inflation factor increases with decreasing $R_o$ (so you might think that doing away with *f* altogether would yield an infinite amount of inflation), it is also true that the time frame during which the inflation occurs simultaneously shrinks to zero. Thus, as the modification to the commutator relation vanishes, so does the inflation.

In Case *c*, we expand the exponential to first order since times are fairly late and find the solution

$$a(t) = a_o'' \exp\left[ -\frac{m}{4}\sqrt{\alpha a^* t} \right] \tag{5.3}$$

which shuts down smoothly as expected. The rate of decrease is mildly dependent on the mass of the field but is still dominated by a*. In this regime, $\rho_c$ begins to take over.

Taking the end of Case *b* as roughly the end of the inflationary period, we can estimate the value of the scalar field according to equation 2.12 by simply plugging in the value of $\eta$. It is quite prodigious.

$$\phi \sim \frac{\eta}{\alpha} \exp\left[ -\frac{\alpha}{2\eta^2} \right] \sim \frac{1}{m^{1/3}} \tag{5.4}$$

where m is the mass of the scalar field in units of Planck masses. If we assume $\phi$ is the Higgs field and take, say, 1 TeV as a reasonable upper limit on its mass, $\phi$ ends up at the end of the inflationary period very far away from anything resembling a low temperature equilibrium value. Furthermore, during most of the inflationary period, $\phi$ has behaved essentially as a free particle. Any interaction term in the Klein–Gordon equation would be negligibly small due to the presence of *f* elsewhere:

$$\nabla_\mu(\hbar f \nabla^\mu)\, \phi + \frac{m^2}{\hbar f}\, \phi + V_{int}[\phi] = 0\ . \tag{5.5}$$

Interactions with other fields become significant only at the end of the inflationary period when $\phi$ has already built up to an enormous value which can then be shed through those other interactions as $\phi$ rolls down to the global minimum of its effective potential, thus reheating the universe. The picture is not entirely unlike Linde's (1983, 1985, 1986b, 1987) chaotic inflationary model from this point on and in fact the value 5.4 of $\phi$ is very nearly what Linde requires as the starting point for his picture. Interestingly, this actually works better for low mass fields such as the Higgs than it would for the immoderately large masses usually contemplated for hypothetical inflaton fields. As we have already seen, the extent of the inflation is mass independent so there is no penalty for considering a low mass field.

**6. Summary**

We have hypothesized that there is a curvature dependence in curved space quantum field theory that cannot be deduced from simply extrapolating flat space quantum field theory. We have then investigated one possible consequence of this hypothesis. No other assumptions are needed in order to provide a vast inflation, solving the usual retinue of cosmological problems. We have also managed to automatically rid the theory of the residual cosmological constant, which usual inflationary theories cannot easily do. This was an unexpected bonus from a choice made during the course of renormalization in the effective action. We believe we made the most economical choice on physical grounds. Finally, we have suggested a model in which the universe creates its contents out of its own expansion, based on the peculiar behavior of the field mode solutions. Furthermore, since the field can be taken to be a real field, such as the Higgs particle, and not some tailor made inflaton field with no other purpose, it is not necessary to concoct elaborate schemes to get rid of the particles so that they are not observed in the current epoch.

There are several obvious areas for further investigation. One of the most important questions addressed by inflationary models is the origin of structure in the universe through the magnification of primordial density fluctuations. Current theories are, however, hard pressed to account for the extremely large structures recently observed (such as the so called "Great Wall"). Perhaps a version of our model will be more accommodating since $f$ dependent fluctuations should be spread over a much broader region than is currently contemplated. The survival of all the normal modes of the field may also play a role here.

Another question which should be addressed is the effect of *f* on higher spin fields. Can inflation be provided by any field of any spin? This seems likely as it arises solely out of the uncertainty principle and would indeed be an advance since inflation has heretofore been tied to a scalar field or something which can be formally treated as such. It would also imply that the vacuum contribution of all fields to the cosmological constant vanishes, thus possibly solving this long–standing problem.

The theory in its present form, however, is only qualitatively realistic since it is based on an unlikely choice for *f*. Again, this choice was made on two grounds: it behaves in a manner qualitatively similar to a more realistic choice (and so we have some confidence in the general tone of the results) and it is analytically soluble. We believe the essential physical content will be unchanged by a more realistic choice but we wished to avoid for now the complications of a numerical analysis. Consequently, this paper should be viewed as a feasibility study, outlining the general shape of the physical content and determining what technical problems arise. There seem to be no insurmountable problems and the results are sufficiently interesting that further study of a more realistic model is warranted.